\documentclass[preprint2]{emulateapj} 

\shorttitle{Approach to Chaos in Planetary Systems} 
\shortauthors{Batygin \& Morbidelli} 

\begin{document}
 
\title{Onset of Secular Chaos in Planetary Systems: Period Doubling \& Strange Attractors}  
\author{Konstantin Batygin$^{1,2}$ \& Alessandro Morbidelli$^2$} 

\affil{$^1$Division of Geological and Planetary Sciences, California Institute of Technology, Pasadena, CA 91125} 
\affil{$^2$Departement Cassiop$\mathrm{\acute{e}}$e: Universite de Nice-Sophia Antipolis, Observatoire de la C$\mathrm{\hat{o}}$te dÕAzur, CNRS 4, 06304 Nice, France}

\email{kbatygin@gps.caltech.edu} 

\begin{abstract}
As a result of resonance overlap, planetary systems can exhibit chaotic motion. Planetary chaos has been studied extensively in the Hamiltonian framework, however, the presence of chaotic motion in systems where dissipative effects are important, has not been thoroughly investigated. Here, we study the onset of stochastic motion in presence of dissipation, in the context of classical perturbation theory, and show that planetary systems approach chaos via a period-doubling route as dissipation is gradually reduced. Furthermore, we demonstrate that chaotic strange attractors can exist in mildly damped systems.  The results presented here are of interest for understanding the early dynamical evolution of chaotic planetary systems, as they may have transitioned to chaos from a quasi-periodic state, dominated by dissipative interactions with the birth nebula.
\end{abstract}

\section{Introduction}
The presence of chaotic motion in planetary systems is well established. As in numerous other dynamical systems, chaos in planetary orbits appears as a result of resonance overlap \citep{1979PhR....52..263C}. For small bodies in the solar system, clustering of various mean-motion resonances leads to chaotic diffusion \citep{1980AJ.....85.1122W}. Indeed, consequences of chaotic motion can be observed both in the asteroid belt, as well as Kuiper belt. The motion of the planets in the outer solar system is also chaotic, but is well bounded, so the system is stable over extremely long periods of time \citep{ 1999Sci...283.1877M}. Conversely, in the inner solar system, secular resonances drive chaotic motion, and the excursions in orbital elements can be quite large \citep{1989Natur.338..237L}. In particular, it has been shown that Mercury's proximity to the $\nu_5$ secular resonance may lead to a dramatic increase in its eccentricity, followed by eventual ejection \citep{1996CeMDA..64..115L, 2008ApJ...683.1207B, 2009Natur.459..817L}. 

All planets in the solar system reside on orbits that are relatively far away from the sun, and as a result, form a nearly undissipative Hamiltonian system. As the discoveries of extra-solar planets have mounted, it has become apparent that a large class of planets reside in close proximity to their host stars. Similarly to the Galilean sattelites system, tidal dissipation plays an important role in the dynamics of these, ``hot" exoplanets. As a direct consequence of tidal dissipation, motion of close-in exoplanets has been assumed to be regular \citep{2002ApJ...564.1024W}. The same can be said for planets and dust particles whose orbital eccentricities and inclinations are constantly damped during early epochs of planet formation. In particular, the presence of gas gives rise to dissipation in the form of stokes drag \citep{1993Icar..103..301B}. Additionally, non-uniform reemission of absorbed  sunlight gives rise to dissipative Poynting-Robertson drag for $\sim \mu$m-sized particles \citep{1982Icar...51..633G}. Thus, little effort has been directed towards the study of chaos, outside of our solar system. In particular, investigations of chaotic motion in planetary orbits, in presence of dissipation, remains a sparsely addressed problem. In this study, we seek to bridge this gap, with an eye towards identification of the dynamical ``route" that planetary systems take between highly dissipated regular motion and chaotic motion within a Hamiltonian framework. It is important to note that this examination has direct astrophysical implications for understanding the dynamical evolution of planetary systems which transition to chaos from a quasi-periodic state that is dominated by dissipative interactions with the nebula, as the gas is slowly removed.

Classical examples of transition from regular to chaotic motion can be found in the context of simple dynamical systems, such as the Logistic Map, which is usually applied to population dynamics, and the Duffing Oscillator, which describes the motion of a forced pendulum in a non-linear potential. In both of the mentioned examples, chaos is approached via the ``period doubling" route, although it is noteworthy that other bifurcations that lead to chaotic motion exist (see for example \citet{2006PhyD..223..194A} and the references therein). In the context of the period doubling approach to chaos, as the degree of dissipation is decreased the periodic orbit, characterized by a period $P$, onto which the system collapses, suddenly changes into a new periodic orbit of period $2P$. When this happens, the periodic orbit transforms into one with two loops, infinitesimally close to each other and to the original shape of the orbit. However, as dissipation is decreased, the twice-periodic nature of the orbit becomes progressively more apparent. If dissipation is decreased further, at some point, the system doubles its period again to $4P$, and so on. As this process is repeated, the period approaches infinity, which is the essence of chaotic motion. In the intermediate regime between a chaotic sea and a $NP$ limit cycle (where $N$ is not too large), resides a dynamically rich structure, known as the ``strange attractor", which is a fractal, possibly chaotic object of intermediate dimensionality. 

In this work, we show that much like the simple examples mentioned above, planetary systems also can approach chaos via the period doubling route. Furthermore, we show that in the context of planetary motion, a strange attractor can exist, given realistic parameter choices which loosely resemble that of the solar system. Our approach to the problem lies in the spirit of classical perturbation theory, where orbit-averaging is employed and only a few relevant terms are retained in the Hamiltonian. In addition to yielding deeper insight into the physical processes at play, this approach is necessary for an efficient exploration of parameter space. Indeed, despite the considerable advances in computational technology in the recent decades, direct numerical integration of dissipative systems remains considerably slower than that of Hamiltonian systems (because symplectic mappings cannot be used), rendering parameter exploration a computationally expensive venture. We begin with a brief review of integrable, linear secular theory for a coplanar planetary system, accounting for eccentricity dissipation in section 2. In section 3, we extend our analysis to non-linear secular perturbations and demonstrate the appearance of global chaos in a purely Hamiltonian framework. In section 4, we add dissipative effects and show the period doubling approach to chaos, and the existence of the strange attractor. We discuss our results and conclude in section 5. 

\section{Linear Secular Theory}

Consider the orbit-averaged motion of a test particle, forced by an eccentric, precessing, exterior planet. We can envision the orbital precession of the planet to be a consequence of perturbations from yet another companion(s), which is too distant to have an appreciable effect on the test particle under consideration. For the sake of simplicity, let us fix the precession rate, $g$, and the eccentricity, $e_p$, of the perturbing planet to be constant in time. If the test particle is far away from any mean-motion commensurability with its perturber, we can write its secular Hamiltonian as
\begin{equation}
H_{sec} = n a^2 \left[ \frac{1}{2} \eta (h^2 + k^2) + \frac{1}{4} \beta  (h^4 + k^4) + \gamma (h h_p + k k_p)  \right],
\end{equation}
where $n$ is the test particle's mean motion, $a$ is its semi-major axis, $h = e \sin \varpi$ and $k = e \cos \varpi$ are the components of the eccentricity vectors, and the subscript $p$ denotes the perturbing planet \citep{1999Sci...283.1877M}. In the above Hamiltonian, $\eta$, $\beta$ and $\gamma$ are coefficients that depend on masses and semi-major axes only, and their functional forms are presented in the appendix. In the regime where $\eta$ does not overwhelmingly exceed other parameters, a Hamiltonian of this form is often referred to as the second fundamental model for resonance \citep{1983CeMec..30..197H}, and also describes mean-motion resonances in the planetary context, although the variables take on different meanings. As will be discussed below, the fourth-order term introduces non-linearity into the equations of motion and renders them non-integrable, allowing for the appearance of chaos \citep{2010arXiv1012.3706L}. 

Although the purpose of this study is to investigate the onset of chaos, it is useful to first consider the regular, integrable approximation. Thus, let us neglect the non-linear term (i.e. set $\beta=0$) for the moment. An application of the linear form of the perturbation equations to the Hamiltonian, yields the equations of motion.
\begin{equation}
\frac{d h}{d t} = \frac{1}{n a^2} \frac{\partial H}{ \partial k} \ \ \ \ \ \ \ \ \ \     \frac{d k}{d t} = - \frac{1}{n a^2} \frac{\partial H}{ \partial h}
\end{equation}

\begin{figure}[t]
\includegraphics[width=0.5\textwidth]{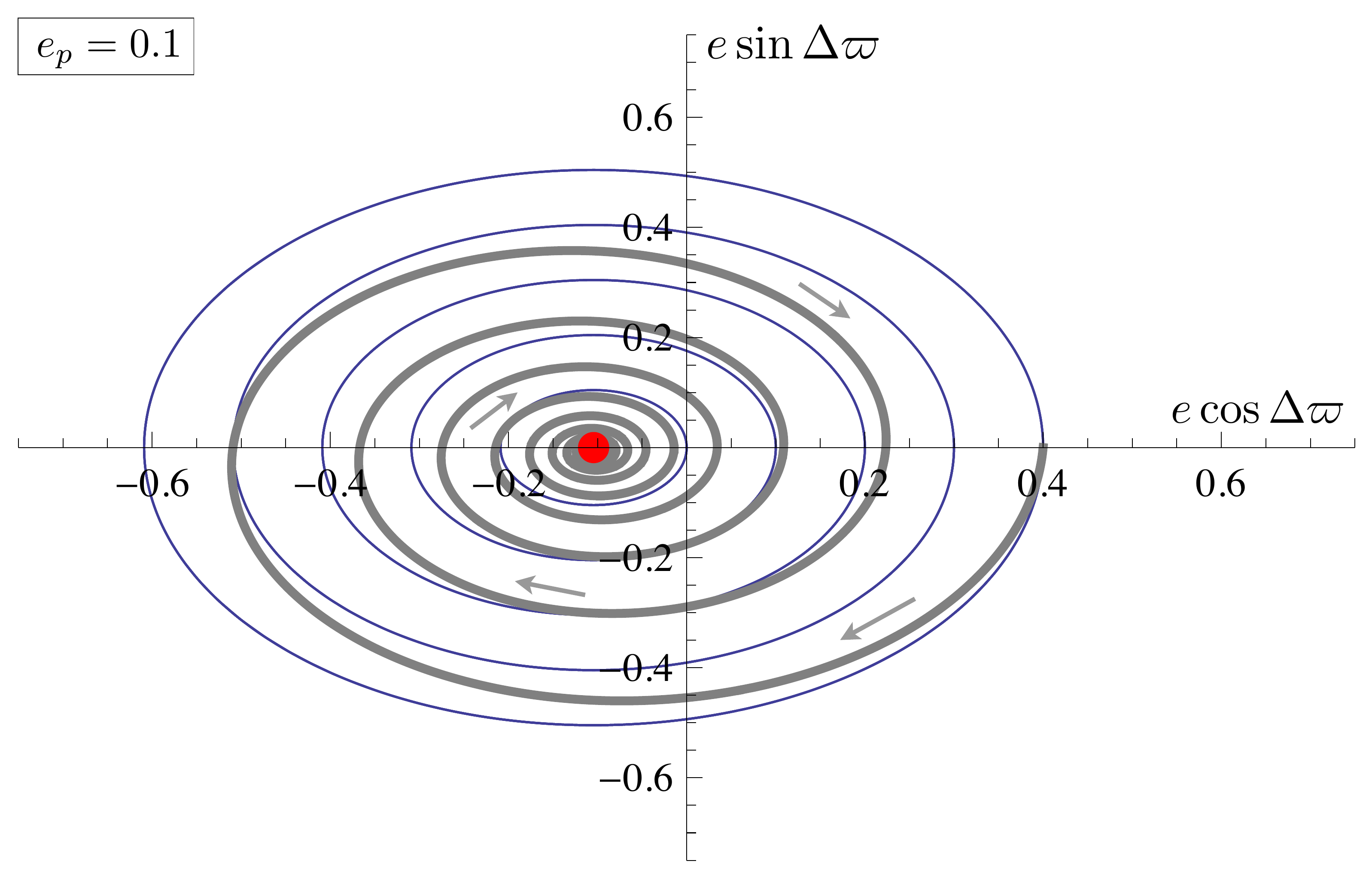}
\caption{Phase space portrait of a test particle, forced by an exterior $m = 15 m_{\oplus}$ perturber, in the linear, integrable approximation. The blue curves depict un-dissipated orbits, while the opaque gray line shows a dissipated orbit, with $\delta = 0.02 \eta$. The red dot onto which the dissipated orbit converges represents the fixed point, which acts as a global attractor for the dissipated system.} 
\end{figure}

Let us now add dissipation into the problem. In planetary systems, dissipation may come about in a number of ways, but is most commonly discussed in the context of tidal friction and interactions of newly formed bodies with a gaseous nebula. Both of these processes lead to a decay of eccentricity and semi-major axes. In the case of tides, the semi-major axes decay time-scale usually greatly exceeds that of the eccentricity, since $\tau_a \equiv a/\dot{a} = e^2 \tau_{circ}$ \citep{1999ssd..book.....M}. Consequently, the decay of semi-major axes can be neglected in most circumstances. The same is generally true for the dissipative effects of the nebula \citep{2002ApJ...567..596L}, although the formalism may be somewhat more complex. Consequently, we model the damping of the eccentricity as an exponential decay with a constant circularization timescale: $ de / dt = - \delta e$, where $\delta = 1/\tau_{circ}$, while we neglect the decay of semi-major axes altogether. Introducing complex Poincar\`e variables, $z = e \exp{i \varpi}$, where $i = \sqrt{-1}$, equations (2) with the inclusion of the dissipative term can be written in a compact form \citep{2002ApJ...564.1024W}:
\begin{equation}
\frac{d z}{dt} = i \eta z + i \gamma e_p \mathbf{e}^{i g t} - \delta z
\end{equation}
This equation of motion admits a stationary periodic solution $z = \gamma z_p / (g -\eta - i \delta)$, which can be expressed as a fixed point in terms of the variable $\tilde{z} = z/z_p$. Note that in $\tilde{z}$, the system (1) becomes autonomous. Physically, this fixed point corresponds to a state where the eccentricity of the particle is constant, while its apsidal line is co-linear and co-precessing with the perturbing planet. Whether the particle is apsidally aligned or anti-aligned with the planet depends on the sign of $(g - \eta)$. 

\begin{figure*}[t]
\includegraphics[width=1\textwidth]{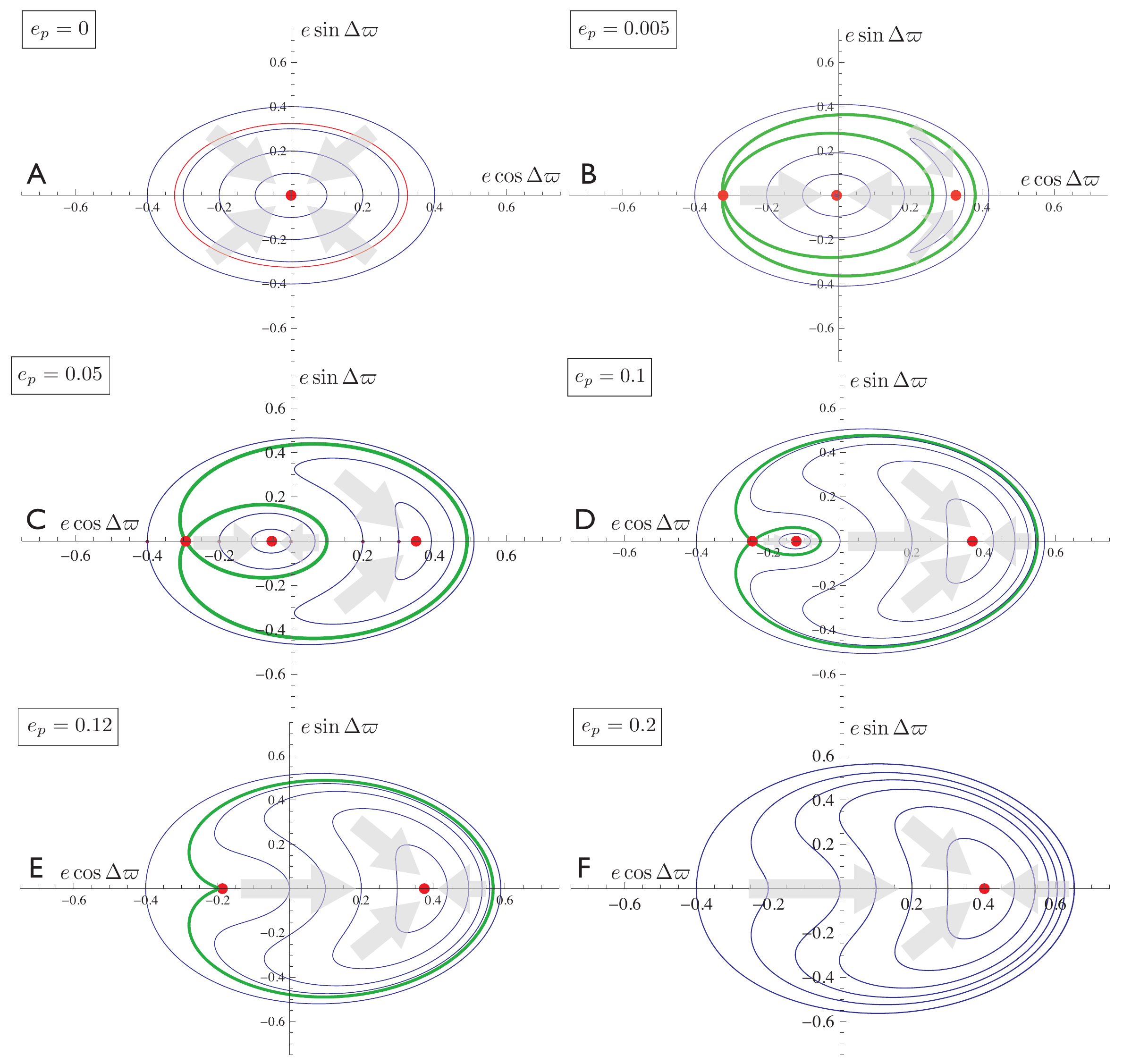}
\caption{Phase space portraits of a test particle, forced by an exterior $m = 15 m_{\oplus}$ perturber, as given by non-linear secular theory. The corresponding perturber eccentricity for each portrait is labeled. Red dots indicate fixed points of a given portrait. The thick green curve, where present, depicts the separatrix. The opaque gray arrows mark the basins of attraction of each stable fixed point.} 
\end{figure*}

The secular fixed point has been discussed in some detail in exoplanet literature. It has been shown that co-planar systems can approach the fixed point on timescales considerably smaller than that of typical planetary system lifetimes, given sufficient tidal dissipation \citep{2002ApJ...564.1024W, 2007MNRAS.382.1768M}. The resulting disappearance of one of the secular eigen-modes from the dynamics of a multi-planet system can yield dynamical stability, where it is otherwise unachievable \citep{2011A&A...528A.112L}. The significantly non-zero inner eccentricity of a fixed point has been invoked to explain ongoing tidal dissipation in close-in planets \citep{2007MNRAS.382.1768M, 2009ApJ...699...23B}. Furthermore, for close-in planets, where tidal precession and general relativity play dominant roles, the exact value of the fixed point eccentricity proves to be a function of the planetary Love number and planetary mass. This allows one to infer information about a transiting planet's interior from its orbit \citep{2009ApJ...704L..49B} and resolve the $\sin(i)$ degeneracy in non-transiting systems \citep{2011ApJ...730...95B}.  

Here, we neglect general relativistic, rotational and tidal precession. This yields perturbation equations that are scale-free (i.e. only dependent on the semi-major axis ratios), but the particular examples shown here are not directly applicable to close-in planets (although the extension of the framework to account for additional precession is very simple - see \citep{2011ApJ...730...95B} and the references therein). 

In the parameter regime described, the general solution to the equation of motion is
\begin{equation}
z = \mathbf{e}^{ (i \eta - \delta)t } \left(c+\frac{ e_p \gamma  \mathbf{e}^{(i g+\delta -i \eta )t}}{g - i \delta -\eta }\right),
\end{equation}
where $c$ is an integration constant that depends on the initial conditions. In absence of dissipation, the phase space portrait is a familiar set of concentric curves that close onto themselves. However, if dissipation is introduced in the system, the phase space area occupied by the orbit begins to contract. Given a sufficient amount of time, the particle settles onto the co-precessing fixed point. This is an important distinction between Hamiltonian and dissipative systems: Hamiltonian flows cannot have attractors \citep{2002mcma.book.....M}. The existence of attractors requires the presence of dissipation.

Figure 1 illustrates a phase space portrait of un-dissipated, as well as damped motion of a test-particle, perturbed by an exterior, $m = 15 m_{\oplus}$ planet, orbiting a Sun-like ($M_{\star} = 1 M_{\odot}$) star. Variables are plotted such that the radial distance depicts the eccentricity of the test-particle, while the polar angle represents the angle between the apsidal lines of the particle and the planet. The blue curves depict un-dissipated orbits, while the opaque gray line shows a dissipated orbit, with $\delta = 0.02 \eta$. The red dot onto which the dissipated orbit converges represents the fixed point, which acts as a global attractor for the dissipated system. The semi-major axes ratio between the test-particle and the planet is chosen to be $\alpha \equiv a/a_p=1/2$, $e_p = 0.1$ and $g = 23"$/year. 

Recall that the position of the fixed point is a function of the perturbing planet's eccentricity. As will be apparent below, this is central to our argument. If the perturbing planet resided on a circular orbit, the fixed point would be at the origin. Furthermore, in our formulation, whether the fixed point is apsidally aligned (to the right of the origin) or anti-aligned (to the left of the origin) depends on the precession rate assigned to the perturbing planet. 

\section{Nonlinear Secular Theory and the Onset of Conservative Chaos}

Now consider the evolution of the test-particle without omitting the fourth-order terms in equation (1). The equation of motion now reads
\begin{equation}
\frac{d z}{dt} = \sqrt{1-|z^2|} \left( i \eta z + i \beta |z^2| z + i \gamma e_p \mathbf{e}^{i g t}\right) - \delta z 
\end{equation}
Note that with $\beta = 0$, and the square root expanded to first order in $e$, we recover equation (3). Here, the square root appears because we no longer limit ourselves to the linear form of Lagrange's planetary equations. The purpose of the square root is to correct for the fact that $(h,k)$ variables are only a low-eccentricity approximation to the true canonical variables, although its inclusion is not instrumental to our results. No general analytical solution for this equation exists, and one must resort to numerical integration to explore the dynamics. As before, it is useful to begin the analysis in absence of dissipative effects.

The addition of the non-linear term introduces important qualitative differences into the solution. First and foremost, if the perturbing planet is eccentric, there are now up to three non-trivial fixed points present, instead of one (e.g. Ch.8 of \cite{1999ssd..book.....M}). One of these fixed points can be unstable (saddle point) and resides on a critical curve (i.e. separatrix) that surrounds, both a librating as well as circulating orbits. Figure 2 shows the phase space portraits of the particle motion, perturbed by a planet of the same parameters as before, but with different eccentricities. 

If the perturber's orbit is circular (Figure 2A), the situation is quite similar to the linear case. In fact, if we omit the square root in equation (5), then a simple analytical solution exists. In this case, the fixed point is at the origin. In direct analogy with the results of the previous section, in presence of dissipation, the fixed point would attract all orbits. There also exists an eccentricity value for the test particle which sets its precession equal to that of the perturber. This set of stationary points is illustrated in Figure 2A as a red circle. However, as long as the perturber's orbit is circular, these stationary configurations are qualitatively no different than any other eccentric orbit. If we now make the perturber slightly eccentric ($e_p = 0.005$), the dynamics changes dramatically (Figure 2B). The first new feature is that the circle of stationary points breaks in two individual fixed points: one unstable at $\Delta \varpi = \pi$ and one stable at $\Delta \varpi = 0$. A critical curve (i.e. the separatrix, green bold curve in panels B-E) is generated at the unstable equilibrium point and encircles the stable one. Second, the stable fixed point that was at the center of the figure moves slightly to the left. 

The appearance of new fixed points has ramifications for dissipative dynamics. As in the linear example, if dissipation (assumed to be finite but much too small to noticeably modify the dynamical portrait, i.e. $\lim \delta \rightarrow 0$) were to be introduced, both of the stable fixed points would act as attractors, with their respective basins of attraction (shown as gray arrows in Figure 2) separated by the critical curve. The stability of fixed points that do not lie on the critical curve, can be understood in the following qualitative manner. Consider a small libration cycle, centered on one of the fixed points. The cycle's intersections with the x-axis are placed symmetrically, relative to the fixed point. The role of dissipation at the higher eccentricity intersection is to decrease the radius of libration, while that at the lower eccentricity intersection is to increase the radius of libration. Of the two antagonist effects, the first wins, because $\dot e \propto e$. Thus, the fixed points centered on libration cycles are stable foci. 

As the eccentricity of the perturber is increased further to $e_p = 0.05$ (Figure 2C) and then to $e_p = 0.1$ (Figure 2D) the phase-space area engulfed by the inner branch of the separatrix (i.e. orbits centered around the stable anti-aligned fixed point) shrinks. Simultaneously, the phase-space area occupied by orbits that are librating around the aligned fixed point grows. When the perturber eccentricity reaches $e_p = 0.12$, the apsidally anti-aligned fixed points collapse onto a single, unstable fixed point (Figure 2E). This implies that if dissipation was to be increased, no apsidally anti-aligned attractor would exist. If the eccentricity of the perturber is enhanced beyond $e_p > 0.12$, the anti-aligned fixed point disappears completely from the portrait (Figure 2D).

In both ``end-member" scenarios we considered ($e_p = 0$ and $e_p = 0.2$), in presence of dissipation, only a single attractor would exist. However, the attractors in these two cases arise from \textit{different} fixed points (i.e. one can not be transformed into another by a change in $e_p$), centered around different branches of the separatrix. Recall that the sole fixed point that was present in the $e_p = 0$ case, disappeared, when $e_p = 0.12$. Similarly, The fixed point that is present in the  $e_p = 0.2$ portrait is not present when the perturber's orbit is circular. This has important implications for the motion of the particle when eccentricity of the perturber is not maintained at a constant value.

\begin{figure}[t]
\includegraphics[width=0.5\textwidth]{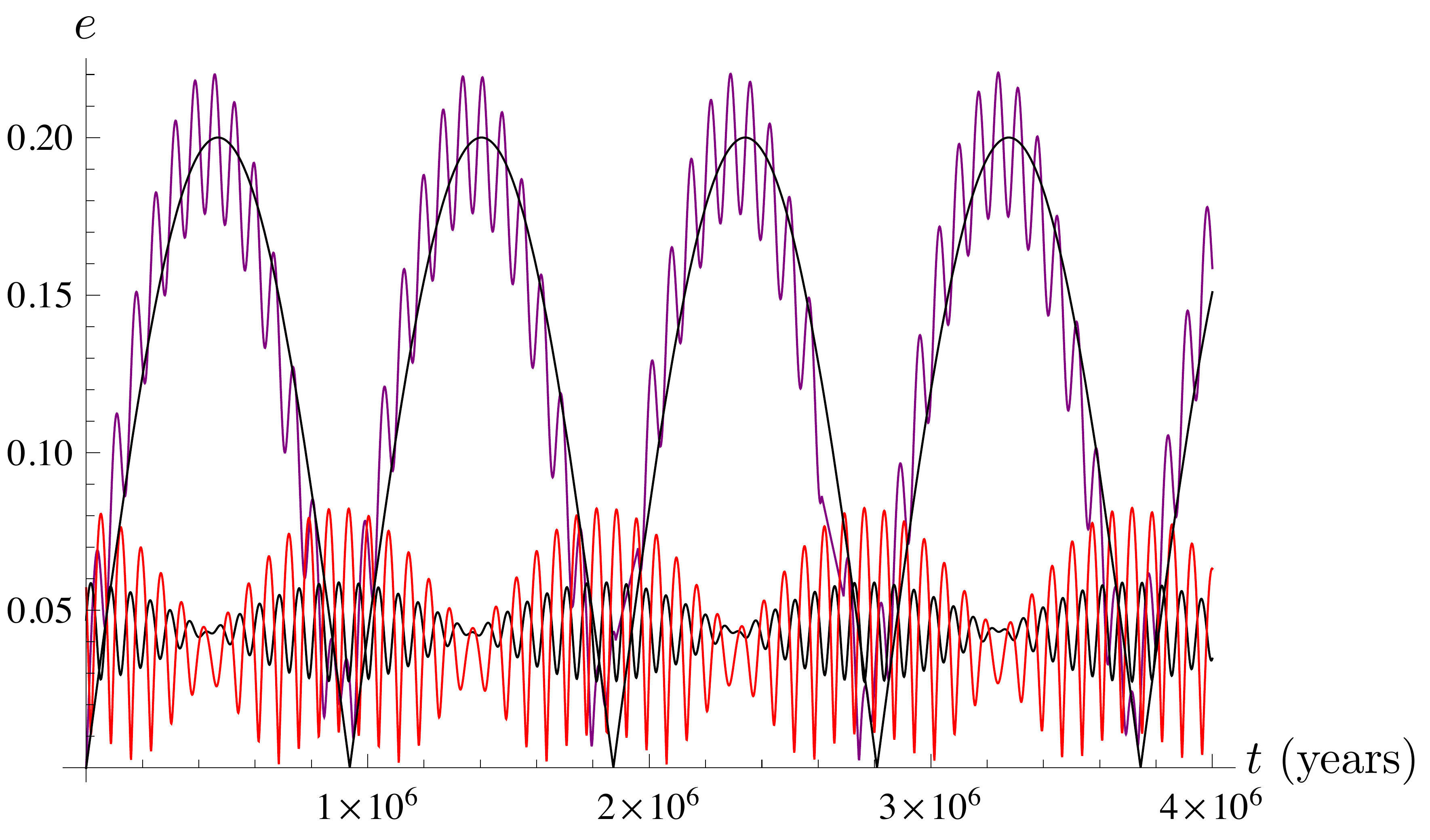}
\caption{Laplace-Lagrange secular solution of Jupiter, Saturn, and a $m = 15 m_{\oplus}$ perturber at $a = 1.85$ AU. The black and red lines, which never go above $e = 0.1$ show the eccentricities of Jupiter and Saturn respectively. The blue and black curves, which attain high eccentricity, represent the exact Laplace-Lagrange solution and the approximate solution, given by equation (6) for the perturber's eccentricity.} 
\end{figure}

Consider a scenario where no dissipation is applied, but the eccentricity of the perturber is varied adiabatically between $e_p = 0$ and $e_p = 0.2$. Here, ``adiabatically" means that the oscillation period of the perturber's eccentricity greatly exceeds the apsidal circulation/libration period of the test-particle. Regardless of the particle's starting condition, its orbit will eventually encounter the separatrix. Since the separatrix is an orbit with an infinite preriod, its crossing necessarily leads to chaotic motion \citep{1989PhyD...40..265B}. In fact the situation is analogous to the motion of an amplitude-modulated pendulum. It has been shown that the chaotic region of such a system occupies the phase-space area that is swept by the separatrix \citep{1991PhyD...54..135H, 1993PhyD...68..187H}. As a result, by ensuring that the eccentricity of the perturber reaches zero and extends above $e_p = 0.12$ at every oscillation, we enforce the entire phase-space within $e \lesssim 0.6$ to be swept by the critical curve, causing all test-particles within this $e$-limit to become chaotic.

Large variations in the perturber's orbital eccentricity can be induced by secular interactions with a distant pair of planets. Consider placing the perturber described above at $a = 1.85$AU, initially on a circular orbit, in presence of Jupiter and Saturn, whose initial conditions correspond to their actual orbits in 1983 (see Ch. 7 of \cite{1999ssd..book.....M}). The orbital evolution of the massive system can be computed using Laplace-Lagrange secular theory, and the resulting solution is presented in Figure 3. Note that this solution is approximate at large eccentricity, since high order terms are neglected. Due to the perturber's proximity to the $\nu_6$ secular resonance, its orbital eccentricity undergoes excursions between $e_p = 0$ and $e_p \approx 0.2$ on a $\tau \approx 2 \pi / (u_p - u_s) \approx 1$ Myr time-scale, where $u$'s are the corresponding eigen-frequencies of the perturbation matrix and $s$ refers to Saturn. Furthermore, the perturber's longitude of perihelion in this solution precesses at a nearly constant rate of $g = 23"$/year. For our purposes we approximate the variation in the perturber's eccentricity as
\begin{equation}
e_p \approx 0.2 | \sin(\frac{u_p - u_s}{2})t | 
\end{equation}

Addition of Jupiter and Saturn to the system does not significantly modify the evolution of the test-particle because of the substantial orbital separation between them ($\alpha_j \sim 0.2$ and $\alpha_s \sim 0.1$). In fact, evaluation of the corresponding constants $\eta$, $\beta$ and $\gamma$ shows that the particle's  interactions with Saturn can be neglected all together, as they only contribute at the $\sim 1\%$ level, while for Jupiter, it suffices to account only for the additional apsidal precession, to which it contributes at the $\sim 30\%$ level, compared to the effect of the considered planet ($15 m_{\oplus}$). Quantitatively, this corresponds to an enhancement of the coefficients $\eta$ and $\beta$, but not $\gamma$. As stated in the Appendix, where the expressions for the constants are given, we have been implicitly retaining the apsidal contribution due to Jupiter since the beginning of the paper, for consistency of the phase-space portraits. The difference in longitude of perihelia between the test-particle and Jupiter forms a comparatively fast angle, and thus can be averaged out. Consequently, we avoid its introduction into the Hamiltonian. This is further warranted, as the magnitude of the interaction term between Jupiter and the test-particle (i.e. $\gamma_j$) is about an order of magnitude smaller than that of the test-particle and the $m = 15 m_{\oplus}$ perturber\footnote{We could have generated an identical chaotic region by introducing another perturber that precesses slightly slower, and tuning its parameters, such that the interaction coefficients in the Hamiltonian, $\gamma$ are equal (see \citet{1990CeMDA..49..177S}, \citet{2010arXiv1012.3706L}). Such a system would constitute a frequency-modulated pendulum rather than an amplitude-modulated one.}.

\begin{figure}[t]
\includegraphics[width=0.5\textwidth]{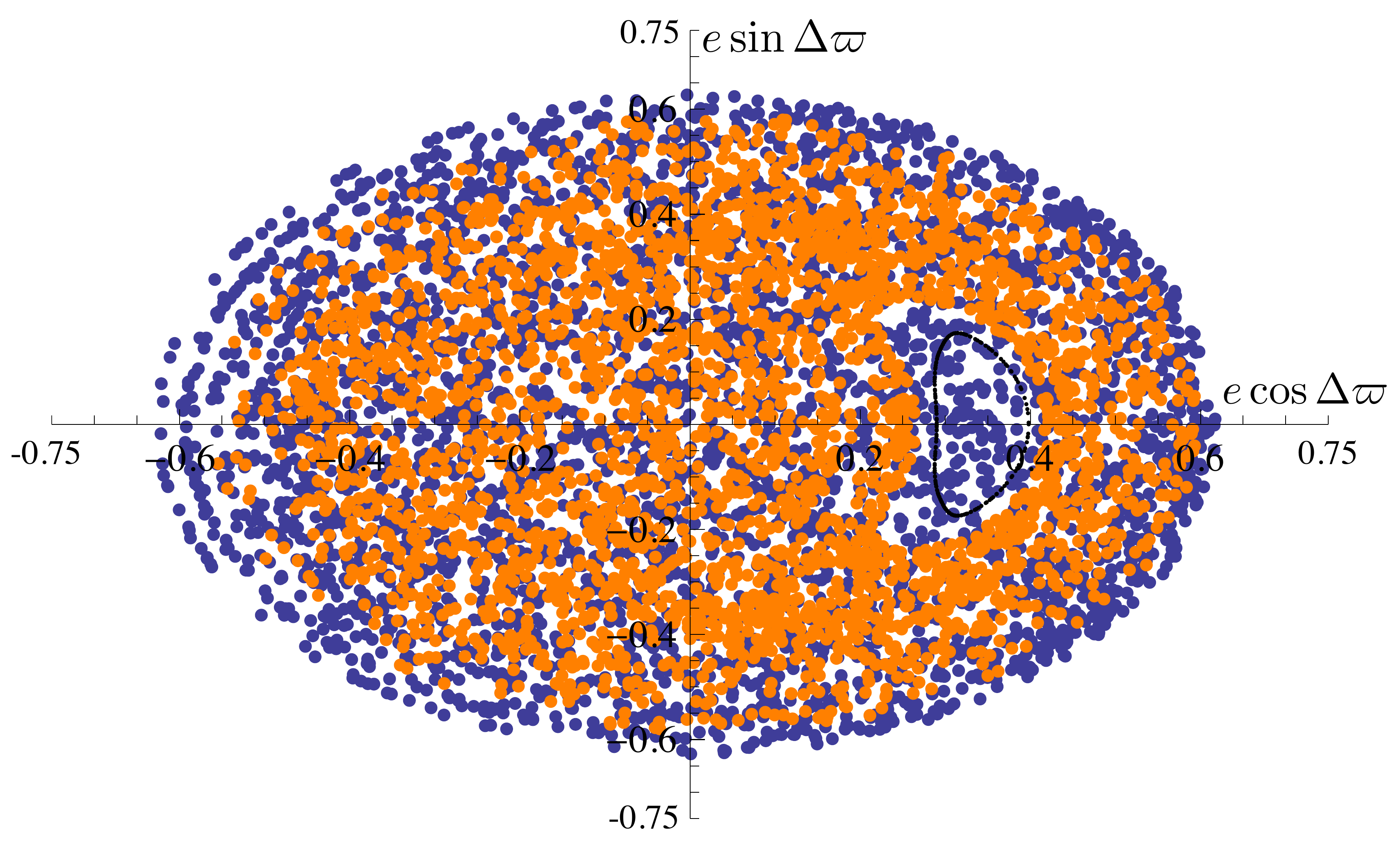}
\caption{Poincar\`e surface of section, illustrating the chaotic dynamics of the test particle. A section is taken at minimum perturber eccentricity. The blue points correspond to an evolution, where the eccentricity of the perturber is given by equation (6). Orange points correspond to an evolution, where the eccentricity of the perturber varies in a similar manner to that described by equation (6), but between $e_p = 0.05$ and $e_p = 0.2$. In the scenario where the perturber's eccentricity does not reach zero, the separatrix fails to sweep the entire phase-space, so a resonant island, roughly outlined by a black orbit, appears.} 
\end{figure}

Since the introduction of the variation of the perturber's eccentricity, we are now faced with a one-and-a-half degrees of freedom Hamiltonian. The dynamics of such a system is best visualized by using a Poincar\`e surface of section. As in the phase-space portraits above, on a Poincar\`e surface of section, a periodic orbit will appear as a point, or a finite sequence of points. A quasi-periodic orbit will appear as a curve that closes upon itself, while a chaotic orbit will appear as a sea of points, which fill a portion of the phase-space. Here, we take a section through phase-space every time $e_p$ goes through zero i.e. with a period of approximately $1$Myr. 

\begin{figure}[t]
\includegraphics[width=0.5\textwidth]{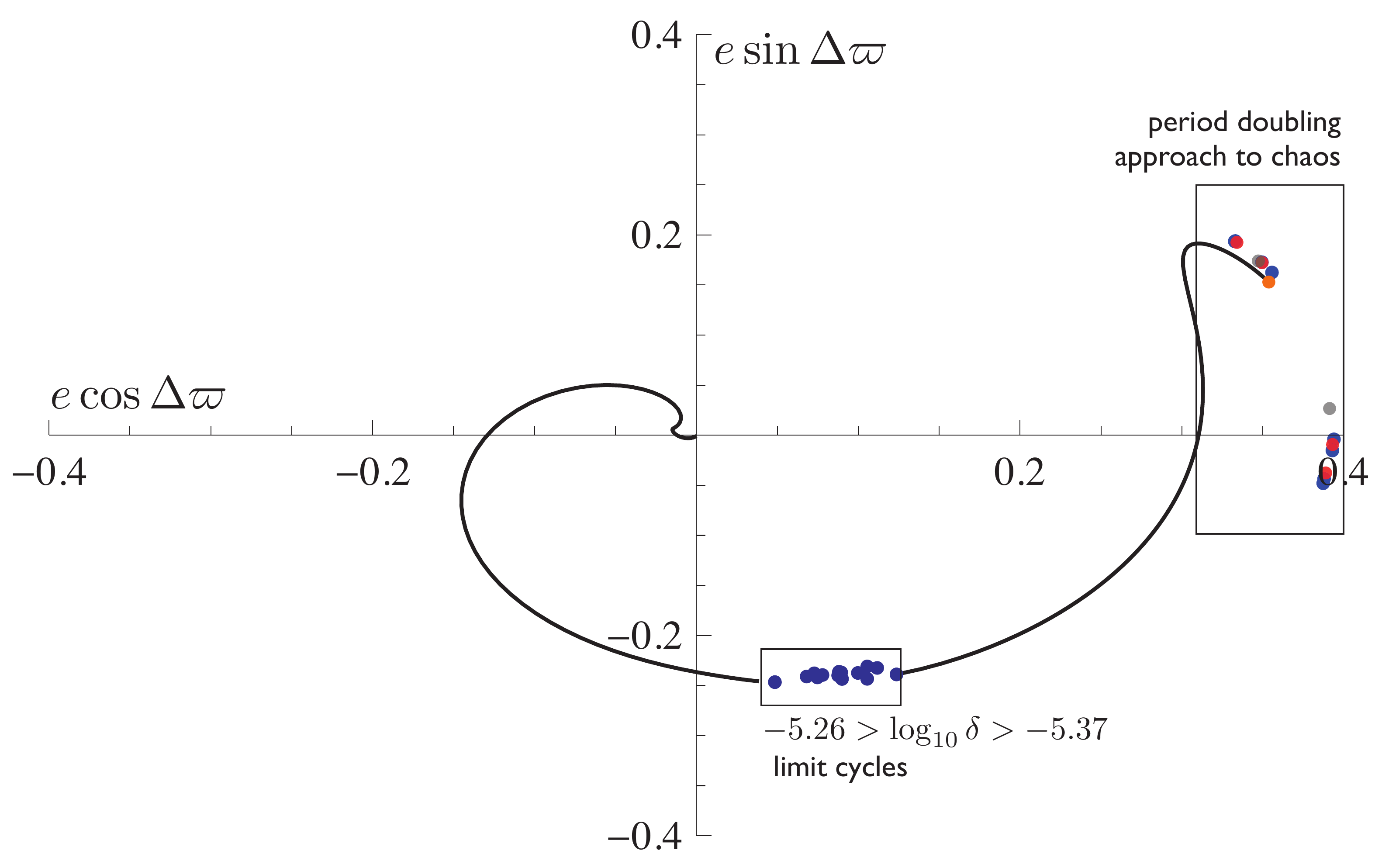}
\caption{This figure depicts the evolution of the fixed point and the subsequent approach to chaos. The black curve shows the movement of the fixed point on a Poincar\`e surface of section. As $\delta$ is reduced, the fixed point leaves the origin and travels outwards in a spiral manner. In the region $-5.26 > \log_{10} \delta > -5.37$, a temporary $2P$ limit cycle is encountered. The period doubling cascade and the onset of chaos (boxed) begins for dissipation rates lower than $\log_{10} \delta = -6.4$.} 
\end{figure}

The Poincar\`e surface of section illustrating the chaotic dynamics of the particle is shown in Figure 4. The blue points correspond to an integration of the system described above, where the eccentricity of the perturbing planet varies according to equation (6). From Figure 4, it is immediately apparent that in this setup, the particle stochastically explores a large fraction of the phase space and no holes appear to exist. We can confirm the chaotic nature of this system by measuring its Lyapunov coefficient, $\lambda$, which is a measure of the exponential divergence rate of nearby orbits:
\begin{equation}
\lambda = \lim_{N \rightarrow \infty} \sum_{i=1}^{N} \frac{\ln(d_{i}/d_{0})}{N \Delta t}
\end{equation}
where $d_0$ is the initial phase-space separation, $d_i$ is the phase-space separation after some time $\Delta t$ and $N$ is the number of renormalizations, where the separation between the orbits is manually returned to the initial value, $d_{0}$ \citep{1976PhRvA..14.2338B}. Adopting $\Delta t$ to be the time between successive sections, $N=500$ and $d_0 = 10^{-6}$, we obtained a positive Lyapunov coefficient of $\lambda = 4.28 \times 10^{-6}$ years$^{-1}$ signifying chaotic motion, with an $\mathbf{e}$-folding timescale of $\tau \sim (g - \eta)/\lambda \sim 7$ secular cycles. Variation of parameters in equation (7) did not change our estimates significantly.

For illustrative purposes, we also performed an integration where the eccentricity of the perturber varies in a similar manner to that described by equation (6), but between $e_p = 0.05$ and $e_p = 0.2$. As already discussed above, in such a scenario, we expect that the particle will not explore the entire phase-space, as the separatrix will fail to sweep all space. The results of this integration are plotted on Figure 4 as orange dots. In accord with the expectations, in this setup, the separatrix falls short of sweeping a considerable section of phase-space and as a result, there exist islands of stability, which the particle never visits. The primary island of stability is outlined by a black orbit in Figure 4 and is centered around the apsidal libration fixed point of the $e_p = 0.05$ phase-space portrait (see Figure 2B). 

\section{Route to Chaos in Presence of Dissipation}

Having constructed a system which exhibits chaotic motion in the previous section, we can now begin to explore the effects of dissipation on chaotic motion. Intuitively, we can expect that in the regime where dissipation dominates all other effects, no chaotic motion can exist. However, the behavior of the orbits in the regime that is intermediate between global chaos and dissipation-dominated motion, is not apparent a-priori.  

\begin{figure}[t]
\includegraphics[width=0.5\textwidth]{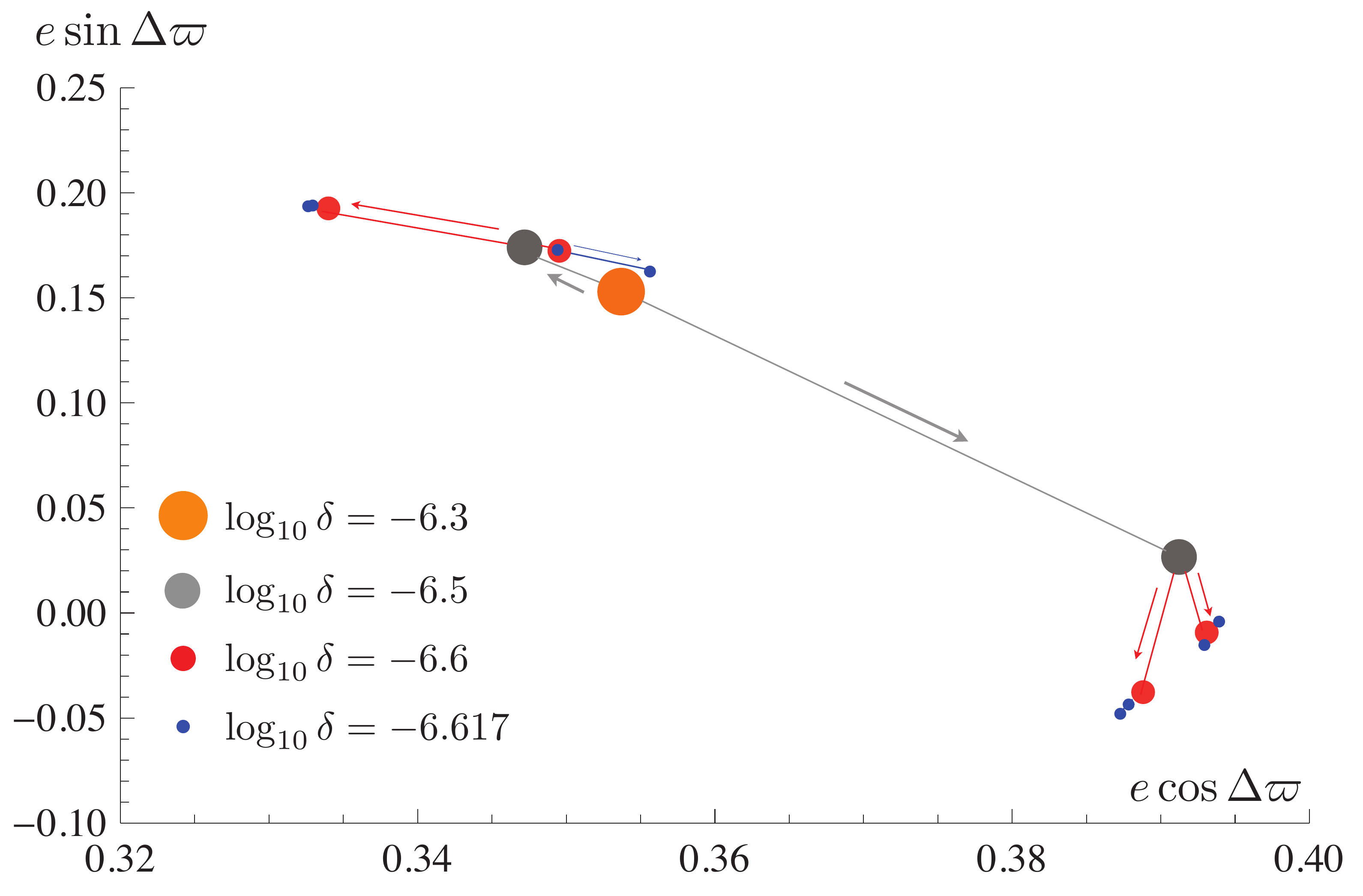}
\caption{Poincar\`e surface of section, showing the period doubling cascade and the approach to chaos. As dissipation is progressively reduced, the fixed point (organge dot), splits into a $2P$ limit cycle (gray dots), which subsequently splits into a $4P$ limit cycle (red dots) and finally into an $8P$ limit cycle (blue dots). Chaos is achieved shortly after.} 
\end{figure}

Our strategy is to begin in the dissipation-dominated regime and track the behavior of the system, while reducing $\delta$ in equation (5). The eccentricity evolution of the perturbing planet is taken to be governed by equation (6). We begin with $\log_{10} \delta = -4$. Numerically, this regime is one where the orbital cirularization timescale, $\tau_c = \delta^{-1}$ exceeds the free precession rate, $\eta$, by a small amount. The solution in this case always falls to a fixed point at the origin. This starting point is optimal, since increasing dissipation further does not change the location of the fixed point onto which the solution collapses. As dissipation is decreased, the location of the fixed point begins to depart from the origin in a spiral manner. This is shown in Figure (5). 

When dissipation is decreased to $\log_{10} \delta = -5.26$, the period of the fixed solution doubles. In other words, the solution falls not onto a fixed point, but onto a limit cycle. A 2$P$ limit cycle appears as two points rather than one on a Poincar\`e surface of section. The series of limit cycles, shown as blue points, corresponding to $-5.26 > \log_{10} \delta > -5.37$ is labeled accordingly on Figure 5. When the dissipation is lowered below $\log_{10} \delta > -5.37$, the orbit once again collapses onto a fixed point. Such behavior is common among systems that approach chaos via period doubling. The trajectory continues to collapse onto a fixed point until $\log_{10} \delta = -6.4$, when the period doubles again, and further decrease in the magnitude of dissipation leads to chaotic motion.

Period doubling is shown in grater detail in Figure (6), which is a zoom-in of the box in Figure 5, labeled ``approach to chaos." At $\log_{10} \delta = -6.3$, the orbit still resides on a fixed point, shown as a large orange dot. At $\log_{10} \delta = -6.5$, the period has doubled and the limit cycle is shown as two gray dots. Decreasing the dissipation further to $\log_{10} \delta = -6.6$, each of the two gray dots splits into two points, giving rise to a 4$P$ limit cycle. This limit cycle is illustrated as four red dots in Figure (6). When dissipation is decreased to $\log_{10} \delta = -6.617$, each of the four red points splits further into two, resulting in an 8$P$ limit cycle. This is shown as series of small blue points in Figure (6). Decreasing the dissipation process repeats the period doubling process. It is noteworthy that once the period doubling process begins, the period of the limit cycle onto which the solution collapses is a very steep function of $\delta$. In other words, the period approaches infinity quickly below $\log_{10} \delta = -6.6$.

As already mentioned in the beginning of the paper, an important feature that dissipative systems can exhibit, which Hamiltonian systems cannot, is the strange attractor. In our setup, the strange attractor appears at $\log_{10} \delta = -6.7$. The phase-space portrait of a strange attractor with $\log_{10} \delta = -6.8$ is illustrated in Figure (7). The strange attractor is in a sense an intermediate state between a limit cycle and global chaos. Although the motion on the attractor itself is chaotic, as can be readily inferred from comparing Figures (4) and (7), it does not occupy the entire available phase space area. This is because the attractor has a diminished dimensionality. Let us consider the dimensionality of the attractors we have encountered thus far. 

Globally chaotic motion, shown in Figure (4) fills the entire available phase space area, and thus its surface of section lies on a two-dimensional manifold. When strong dissipation was introduced into the problem, the motion collapsed onto a fixed point which is zero-dimensional object. Limit-cycles have surfaces of section of dimensionality between 0 and 1. A dimensionality of unity is achieved if the motion collapses onto a limit-torus, whose surface of section appears as a curve that closes upon itself. Further decrease in dissipation results in the appearance of strange attractors, whose surfaces of section lie on manifolds of intermediate dimensionality, between 1 and 2. 

The dimensionality of the phase-space portrait can be related directly to the Lyapunov exponent and thus presence of chaos. If motion that originates from different initial conditions converges onto a single fixed point or limit-cycle attractor, the Lyapunov exponent must be negative, signaling periodic motion. For example, in our system, a fixed point with $\log_{10} \delta = -6$ is characterized by $\lambda = -8.1 \times 10^{-7}$ years$^{-1}$. The sign of the Lyapunov exponent changes if the dimensionality of the attractor exceeds unity. Indeed, motion on the strange attractor, shown in Figure (7) is characterized by $\lambda = 1.87 \times 10^{-6}$ years$^{-1}$. Note that although motion is chaotic, the $\mathbf{e}$-folding timescale corresponds to $\sim 16$ secular cycles, a factor of $\sim 2$ longer than that of the undissipated system.

\begin{figure}[t]
\includegraphics[width=0.5\textwidth]{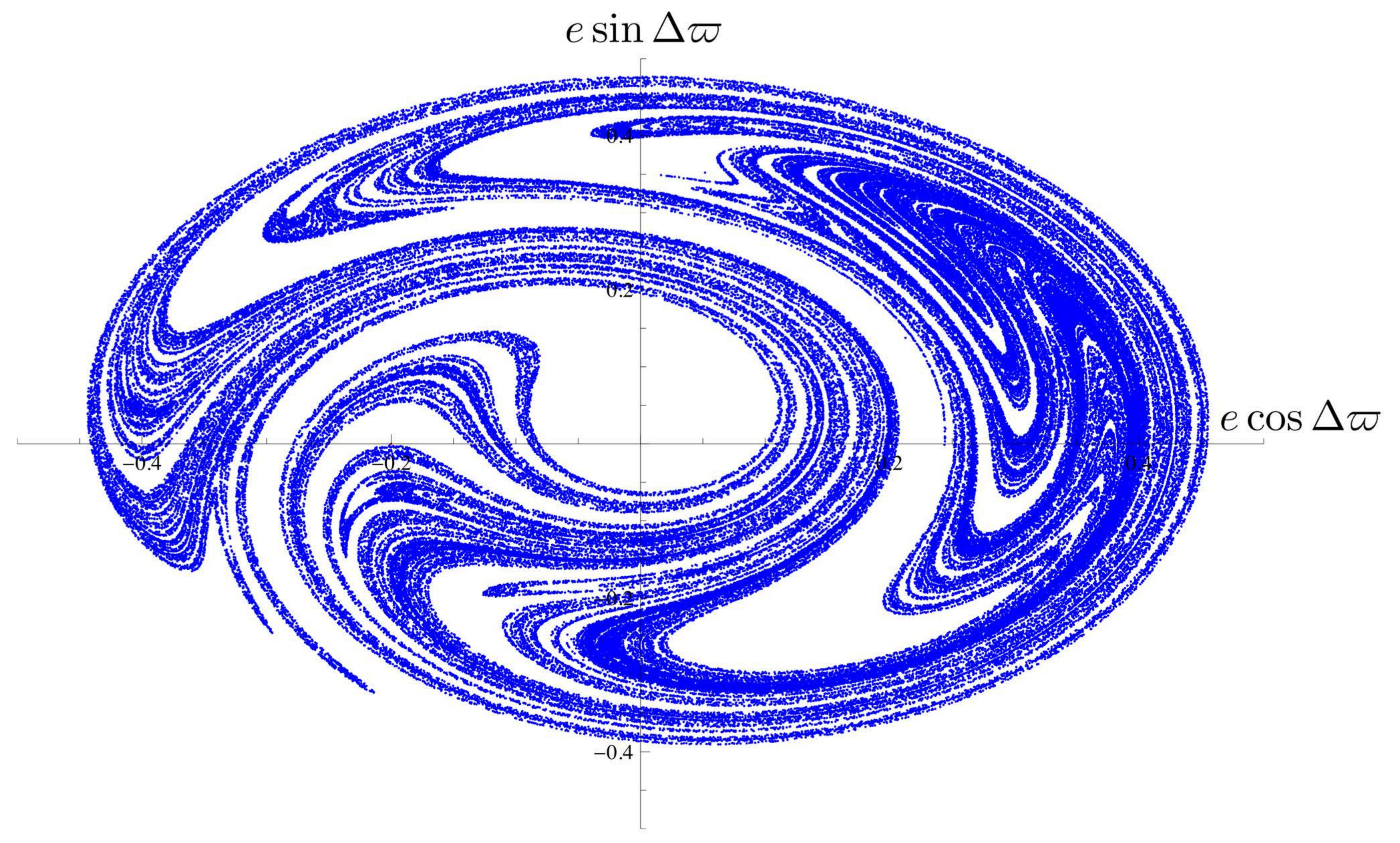}
\caption{A strange attractor. The shown object corresponds to a dissipation level of $\log_{10} \delta = -6.8$ and is characterized by a positive Lyapunov coefficient of $\lambda = 1.87 \times 10^{-6}$, signaling chaotic motion on the attractor. The Minkowski--Bouligand dimensionality of this particular attractor is $D = 1.75 \pm 0.005$. Its persistence requires the lack of islands of stability in the occupied region of phase-space.} 
\end{figure}

There are many ways to define a fractal dimension. Here, we shall work in terms of the Minkowski--Bouligand dimension (see for example \citet{1993cds..book.....O}). The Minkowski--Bouligand dimensionality of a particular strange attractor can be computed by utilizing a \textit{box-counting} algorithm. In this approach, the phase-space is divided into an even number of sub-regions (boxes) and the number of boxes, occupied by the attractor is counted. This is performed over a large number of scales, typically decreasing the box size by a factor of 2 upon each iteration. The slope of the line which describes the number of occupied boxes as a function of the box size in $\log-\log$ space is the dimensionality of the object. We have performed this calculation for the strange attractor presented in Figure (7), covering 6 scales. As a result, we find that the attractor has a dimensionality of $D = 1.75 \pm 0.005$. 

As can be expected from simpler examples, such as the Duffing Oscillator, the strange attractor is only present for a limited range of parameters. We have performed additional experiments where dissipation was decreased further. We found that the majority of the attractor breaks up into large chaotic regions by $\log_{10} \delta = -7.7$, and at $\log_{10} \delta = -8$, the surface of section is once again essentially filled, signaling a return to the global chaotic sea, characteristic of the undissipated system (Figure 4).

In the analysis above, we started from a configuration, where in absence of dissipation, chaos engulfed all available phase-space, in which no holes existed. Consider what would happen if we were to redo the experiment using a slightly different setup, that yields a small island of stability around the apsidally aligned fixed point, as shown in Figure 4. In such a system, the strange attractor would not persist, for eventually, the particle would necessarily end up in the neighborhood of the island of stability, and driven by dissipative effects, find itself in the basin of attraction of the fixed point. This implies that in configurations where chaos is not global, the presence of dissipation tends to guide the constituents towards regular orbits, with the degree of dissipation directly dictating the time-scale needed for the system's arrival to a quasi-periodic state.

\section{Discussion}

In this paper, we investigate the onset of chaotic motion in planetary systems, where dissipative effects play an important role. Using a semi-analytical perturbation approach, we have shown that planetary chaos appears through a period doubling cascade. We have further demonstrated that strange attractors can exist in the context of a planetary problem under the condition of global chaos in absence of dissipation.

It is important to consider the astrophysical significance of this process, beyond purely academic interest. As already mentioned in the beginning of the manuscript, one can expect the period doubling cascade to occur during early epochs of a planetary system's dynamical evolution, as the evaporation of the birth nebula leads to a gradual decrease in dissipation. As a result, the work presented here describes how the dynamical portrait of a system may evolve shortly after formation, when gas is gradually taken away. Note that the dissipation timescale, corresponding to the strange attractor, is typical of a late-stage proto-planetary disk (i.e. $\tau_{circ} \sim 10^6$ years) \citep{2002ApJ...567..596L}. In other words, the example configuration described here may correspond to a planetary system where the planets are already massive enough to have essentially decoupled from the gas, and are forcing a small planetesimal, which still feels considerable drag.

Other forms of dissipation, such as tides, are abundantly present in the planetary context, and are of special importance for hot exoplanets. This is further relevant, since understanding the dynamics of multiple close-in planets is becoming increasingly important, as their numbers in the observed aggregate grow. Particular interest is exhibited towards orbital configurations that converge to a fixed point, since a stationary system is required for obtaining an estimate for the Love number, $k_2$, of extra-solar gas giants \citep{2009ApJ...704L..49B}. Although linear theory predicts that a dissipated system's arrival to a stationary configuration is only a matter or time (i.e. tidal $Q$) \citep{1966Icar....5..375G}, non-linear theory presented here, suggests that one should exercise caution, as a fixed point is not always the end-state.

The same problem can also be turned around. We have shown here that limit cycles reside in limited parameter regimes. Thus, an observed system, whose orbital evolution follows a limit-cycle, can be used to place desperately needed constraints on the tidal quality factor, which remains among the most unconstrained parameters in planetary science and whose physical origin is an area of ongoing research \citep{2005ApJ...635..688W}. 

For close-in planets, the effect of general relativity and tidal precession plays in favor of approach to the fixed point, rather than any other attractors. This is because it enhances the coefficients $\eta$ and $\beta$, but not $\gamma$, in the Hamiltonian. Since the orbital precession of a putative external perturber will generally be comparatively slow, the enhanced precession of close-in planets will tend to de-tune any resonance. As the relative amplitude of the external perturbations is diminished, the dynamics approaches the $e_p = 0$ phase-space portrait seen in Figure 2A. Naturally, this will also lead to stabilization of the orbits. It is noteworthy that such an effect is also at play in the solar system, as additional precession from GR places the free precession of Mercury further from the $\nu_5$ secular resonance, diminishing its chances of ejection \citep{2008ApJ...683.1207B, 2009Natur.459..817L}. 

Finally, it is worthwhile to consider the limitations of the presented model. Indeed, we have approached the problem by utilizing a classical perturbation theory, where only a few relevant terms are retained in the disturbing function. As already mentioned above, this approach was necessary for the exploration of parameter space, as the efficiency offered by conventional direct integration is not sufficient. Although the approach we take here breaks down at high eccentricities, in the solar system, it has been successful in capturing the important physical processes that govern chaotic motion \citep{1990CeMDA..49..177S}. The recent work of \citet{2010arXiv1012.3706L} has further confirmed this to be true in the case of Mercury's orbit. Thus, we expect that inclusion of the full disturbing function will only modify our findings on a quantitative level. However, future numerical confirmation and re-eavluation of the work done here will surely be a fruitful venture, especially if performed in the context of a particular observed system.

\textbf{Acknowledgments} We thank K. Tsiganis for numerous useful discussions and Oded Aharonson for carefully reviewing the manuscript. Additionally, we thank the anonymous referees for useful suggestions.

\begin{appendix}

\subsubsection{Coefficients of the Hamiltonian}
In this work, we choose to write the coefficients, such that they appear in the equations of motion without pre-factors. The notation used here is identical to that of \citet{1999ssd..book.....M}, i.e. $b$ denotes a Laplace coefficient, $\alpha$ is the semi-major axis ratio, and $\mathcal{D} \equiv \partial/\partial \alpha$. Throughout the paper, we account for the induced precession that arises from the $m = 15 m_{\oplus}$, $\alpha_p = 1/2$ perturber as well as Jupiter, with $\alpha_J = 0.178$, but not Saturn. Evaluation of the formulae below shows that Saturn's effect is negligible. The eccentricity forcing, taken into account is solely due to the $m = 15 m_{\oplus}$ perturber. The resulting formulae read:
\begin{equation}
\eta = \frac{n}{4} \left(\frac{m_p}{M_{\star}}\frac{a}{a_p} \left(2 \alpha_p \mathcal{D} + \alpha_p^2 \mathcal{D}^2 \right) b_{\frac{1}{2}}^{(0)}(\alpha_p) + \frac{m_J}{M_{\star}}\frac{a}{a_J} \left( 2 \alpha_J \mathcal{D} + \alpha_J^2 \mathcal{D}^2 \right) b_{\frac{1}{2}}^{(0)}(\alpha_J)  \right)
\end{equation}
\begin{equation}
\beta = \frac{n}{32} 
\left(\frac{m_p}{M_{\star}}\frac{a}{a_p} \left( 4 \alpha_p^3 \mathcal{D}^3 + \alpha_p^4 \mathcal{D}^4 \right) b_{\frac{1}{2}}^{(0)}(\alpha_p) + \frac{m_J}{M_{\star}}\frac{a}{a_J} \left( 4 \alpha_J^3 \mathcal{D}^3 + \alpha_J^4 \mathcal{D}^4 \right) b_{\frac{1}{2}}^{(0)}(\alpha_J)  
\right)
\end{equation}
\begin{equation}
\gamma = \frac{n}{4} \frac{m_p}{M_{\star}}\frac{a}{a_p} \left( 2 - 2 \alpha_p \mathcal{D} - \alpha^2 \mathcal{D}^2 \right)  b_{\frac{1}{2}}^{(1)}(\alpha_p) 
\end{equation}
The first and the second terms in $\eta$ and $\beta$ arise from the $m = 15 m_{\oplus}$ perturber and Jupiter respectively. All terms in $\gamma$ correspond to the $m = 15 m_{\oplus}$ perturber.

\end{appendix}

\end{document}